\documentclass[a4paper,11pt]{article}
\pdfoutput=1 

\usepackage{jheppub} 

\usepackage[T1]{fontenc} 
\usepackage{comment}

\newcommand{\be}{\begin{equation}}
	\newcommand{\ee}{\end{equation}}

\def\a{{\alpha}}
\def\l{{\lambda}}

\def\b{{\beta}}

\def\Q{{\mathcal{Q}}}

\newcommand{\bea}{\begin{eqnarray}}
	\newcommand{\eea}{\end{eqnarray}}
\newcommand{\ba}{\begin{eqnarray}}
	\newcommand{\ea}{\end{eqnarray}}
\newcommand{\nn}{\nonumber \\}

\title{Inconsistency of point-particle dynamics on  higher-spin backgrounds}

\author[a,b]{Vyacheslav Ivanovskiy}
\author[b,c]{and Dmitry Ponomarev}

\affiliation[a]{Moscow Institute of Physics and Technology,
  Dolgoprudny, 141701, Russia}
\affiliation[b]{Institute for Theoretical and Mathematical Physics,\\
Lomonosov Moscow State University,  Moscow, 119991, Russia}
\affiliation[c]{I.E. Tamm Theory Department, Lebedev Physical Institute,
 Moscow, 119991, Russia}

\emailAdd{ivanovskiy.va@phystech.edu}
\emailAdd{ponomarev@lpi.ru}

\abstract{We use the light-cone gauge formalism to study interactions of point particles with massless higher-spin fields. By analysing the light-cone consistency conditions at the subleading order in higher-spin fields, we find that no local interactions of point particles with chiral higher-spin fields are possible.
Considering that chiral higher-spin theories form inevitable closed subsectors of any consistent massless higher-spin theories in flat space, this conclusion holds more generally, in particular, it applies to putative parity-invariant completions of chiral higher-spin theories. 
 Besides that, we argue that our result implies that Riemannian geometry cannot be extended to spaces with non-trivial higher-spin fields, in particular, there is no  higher-spin extension of space-time interval. In the present paper we focus on a case of a massless particle, while a more technical massive case will be analysed in a companion paper.}

\begin{document} 
\maketitle
\flushbottom

\section{Introduction}

Higher-spin theories are theories of massless fields with spin two and higher. The spin-two sector of a theory is supposed to be identified with gravity, probably, in a modified form.
Each massless higher-spin field comes together with the  associated symmetry. Accordingly, higher-spin theories can be viewed as highly symmetric extensions of gravity. Considering the important role symmetries play in controlling ultraviolet divergences, one  expects, that the extension of gravity with higher-spin fields may improve its quantum properties, thus, giving prospective models of quantum gravity. At the same time, the presence of higher-spin symmetries also results in substantial difficulties. Namely, higher-spin symmetries appear to be so extensive, that they complicate the construction of higher-spin theories themselves. These difficulties are showcased by a broad list of  no-go theorems, stating that under some rather general and natural sets of assumptions massless higher-spin fields cannot interact, see e. g. \cite{Weinberg:1964ew,Coleman:1967ad,Aragone:1979hx,Maldacena:2011jn,Sleight:2017pcz,Sleight:2021iix}.  Nevertheless, there are few special setups, which require relaxing one or another standard assumption of these theorems, for which higher-spin theories do exist. Recent reviews on higher-spin theories can be found in \cite{Bekaert:2010hw,Didenko:2014dwa,Vasiliev:2014vwa,Bekaert:2022poo,Ponomarev:2022vjb}.

In the present paper, we will deal with a particular class of higher-spin models, known as chiral higher-spin theories  \cite{Metsaev:1991mt,Metsaev:1991nb,Ponomarev:2016lrm}.
 Chiral higher-spin theories  are manifestly local massless higher-spin theories in four-dimensional flat space  and they admit a concise formulation at the level of action. This is achieved at the expense of relaxing parity.  More specifically, chiral higher-spin theories can be regarded as higher-spin extensions of self-dual Yang-Mills theory and self-dual gravity \cite{Ponomarev:2017nrr,Krasnov:2021nsq}\footnote{A different theory that involves interactions of higher-spin fields and relies on the same mechanism was suggested in \cite{Devchand:1996gv}.}.
  Self-duality entails a list of properties, which chiral higher-spin theories posses. In particular, these are integrable and, as a result, the associated tree-level scattering is trivial.
 Besides that, as for the lower-spin self-dual theories, the action of the chiral higher-spin theory is not real in the Lorentzian signature.
 Due to these and other features, chiral higher-spin theories can hardly be regarded as a completely satisfactory solution to the higher-spin problem and it would be important to extend these away from the self-dual sector.
 
  At the same time, while such a parity-invariant extension is not known, we can explore what is already available to us. As lower-spin examples showcase, self-dual theories and their parity-invariant counterparts share many common structures, such as symmetries, classical solutions and amplitudes.
 In a similar manner, chiral higher-spin theories form unavoidable closed subsectors of their putative parity-invariant completions and can be used to access certain observables and study the properties of the latter, although these are not available. In particular, one can study various couplings to higher-spin fields --  as we will do below -- which are severely constrained already in the self-dual sector. 
 
 Since the chiral higher-spin theory was suggested, it was explored and extended in different ways. In particular, the underlying color-kinematics duality structure was established \cite{Ponomarev:2017nrr,Monteiro:2022lwm,Monteiro:2022xwq,Ponomarev:2024jyg}, one-loop corrections for chiral higher-spin amplitudes were computed \cite{Skvortsov:2018jea,Skvortsov:2020wtf}, chiral higher-spin theories were covariantised and deformed to anti-de Sitter space \cite{Krasnov:2021nsq,Sharapov:2022faa,Sharapov:2022wpz}, where they can be embedded  \cite{Didenko:2022qga} into the framework of  \cite{Vasiliev:1990en}. Besides that, twistor-space formulations of chiral higher-spin theories have been explored \cite{Herfray:2022prf,Adamo:2022lah,Neiman:2024vit}, holographic duality involving chiral higher-spin theories was suggested \cite{Ponomarev:2022ryp,Ponomarev:2022qkx}, an extension to six dimensions was proposed \cite{Basile:2024raj}, as well as some classical solutions were found \cite{Skvortsov:2024rng,Tran:2025yzd}. In the present paper we will study yet another aspect of chiral higher-spin theories. Namely, aiming at better understanding of higher-spin geometry, we will study couplings of point particles to chiral higher-spin theories.

In the same way as higher-spin symmetries may help improving ultraviolet properties of gravity, one may expect that the presence of higher-spin fields may also help resolving some problems of classical General Relativity, in particular, difficulties related to singularities. The analysis of singularities and, more generally, of any geometric features of classical higher-spin solutions, however, quickly encounters a serious issue. Namely, as it is not hard to see, the familiar geometric notions associated with General Relativity prove to be not invariant with respect to higher-spin symmetries, see e. g. \cite{Didenko:2009td,Ammon:2011nk,Iazeolla:2011cb,Castro:2011fm,Vasiliev:2014vwa}. In particular, the position of a horizon of a black-hole-like higher-spin solution can be altered by higher-spin gauge transformations. This, clearly, indicates that the standard geometric tools from General Relativity are not directly applicable to the case when higher-spin fields and symmetries are present.

There is, however, a possibility that these can be properly amended with higher-spin dependent corrections in such a way that the resulting higher-spin geometric notions are, indeed, compatible with higher-spin symmetries. This is the problem, which we will address below. More precisely, we will study whether scalar point particles can be consistently coupled to chiral higher-spin theories. Considering that all the familiar geometric notions are derivable from point-particle motion in space-times the geometry of which is being studied -- for example, the space-time interval is just the on-shell action of a scalar point particle -- by coupling a point particle to higher-spin fields, we will be able to extend the standard geometric constructions to the higher-spin case\footnote{In the higher-spin context ''geometry'' can be understood in many different ways. In particular, for free theories \cite{deWit:1979sib} introduced generalised higher-spin curvatures and higher-spin Christoffel-like connections.  Accordingly, the approach to higher-spin theories, that relies on these notions was referred to as geometric in \cite{Francia:2002aa,Francia:2002pt}. The frame-like formalism of \cite{Vasiliev:1990en} employs connection-like one-forms, generalising the associated ingredients of the Yang-Mills theory and of the Cartan-Weyl approach to gravity and, in this sense, it can also be regarded as geometric. In the present paper we understand ''geometry'' in the most standard sense, that is as a set of tools that allows one to define lengths, angles, volumes, parallel transport and curvature, etc. All these can be derived in the usual manner in terms of the dynamics of scalar point particles in the given space-time. One can envisage other versions of geometry, based on the dynamics of other probe objects. Some of these alternatives as well as the existence of their coupling to higher-spin theories are briefly discussed in section \ref{sec:6}.}.  

It is important to mention, that a consistent coupling of point particles to higher-spin fields is known already \cite{Segal:2000ke,Segal:2001di}. Symmetries of this coupling eventually lead to conformal higher-spin theory \cite{Segal:2002gd}\footnote{It needs to be clarified, that 
 the construction based on the point-particle dynamics initially leads to a theory,  which does not involve space-time derivatives in the action. To obtain a theory with meaningful dynamics it needs to be deformed. After the deformation, the resulting conformal higher-spin theory does not couple to a point particle any more, instead, it couples to a conformal scalar field. In the latter form, conformal higher-spin theory was also obtained    \cite{Tseytlin:2002gz} as a one-loop effective action of a conformal scalar field in a higher-spin background.}. The main drawback of this construction is that conformal higher-spin fields, unlike massless ones, are not unitary. Few attempts to construct a massless higher-spin theory building up on the setup \cite{Segal:2000ke,Segal:2001di} were undertaken, see e. g.  \cite{Segal:2001qq,Ponomarev:2013mqa}, but these were not successful.
Similarly, attempts to use the constructions from Finsler geometry -- an extension of  Riemannian geometry with higher-rank tensors -- so far failed to provide actions, which would respect symmetries of massless higher-spin fields, see \cite{Tomasiello:2024jyu} and references therein. In other words, although some interesting suggestions are available in the literature, the problem of coupling of point particles to massless higher-spin fields remains open. Below we will systematically revisit this issue having a concrete massless higher-spin theory known from the beginning\footnote{Interactions of point particles and higher-spin fields were also discussed from other perspectives, see e. g. \cite{Bastianelli:2007pv,Grigoriev:2021bes}.}.

First results on the coupling of a point particle to massless higher-spin fields were obtained in \cite{deWit:1979sib}, where the point particle action was found at the leading order in fields.  In the unfolded approach the leading order coupling  was obtained  in \cite{Tarusov:2023rad}. In a recent work  \cite{Ivanovskiy:2023aay}, we rederived this result in the light-cone formalism.  In the present paper we will use this formalism to  study interactions of point particles with chiral higher-spin fields at the second order in interactions. At this order one expects to find stringent constraints, which either fix the interactions up to a very limited set of coupling constants or rule them out. Focusing on the case of massless point particles, we will show that they cannot consistently interact with chiral higher-spin fields and, as a consequence, with any massless higher-spin theory in four-dimensional Minkowski space.

This paper is organised as follows. In section \ref{sec:2} we review  the light-cone formalism for free massless fields and free scalar point particles in four-dimensional Minkowski space. We then proceed to the non-linear level in section \ref{sec:3}, where we review the light-cone consistency conditions on interactions, as well as give the necessary prerequisites on chiral higher-spin theories. In the following sections we analyse these consistency conditions. First, we solve the kinematical constraints in section \ref{sec:4} and then analyse the dynamical constraints at two leading orders  in section \ref{sec:5}.
We conclude in section \ref{sec:6}. Two appendices summarise our notations and contain some technical results.

\section{Free theories}
\label{sec:2}

In our analysis we will use the light-cone formalism. This approach was used  to construct  chiral higher-spin theories originally and it is particularly suitable for dealing with them.  
Unlike in  covariant approaches, in the light-cone formalism the gauge freedom is fixed from the beginning. Due to that, the analysis of consistent interactions of point particles with massless fields in the two approaches works very differently. Namely, in the covariant approach, consistency of a coupling of a point particle to a particular gauge theory translates into the requirement that the point particle action is invariant with respect to the gauge symmetries of the field theory. Instead, in the light-cone formalism, the key consistency condition is the requirement that the action of a point particle coupled to massless fields is Poincare invariant. The relevant information about a particular massless theory that contributes to this analysis is not its gauge symmetries, but the transformation laws of gauge-fixed fields under the Poincare algebra action. 

The light-cone formalism is a systematic approach that allows one to construct interacting Poincare-invariant theories in the light-cone gauge perturbatively. To control Poincare symmetry one explicitly constructs the phase space Noether charges associated with each generator and requires that these commute properly. 
	In this section  we briefly review the light-cone formalism for the free massless fields and the free point particle in  Minkowski space. We then proceed to interactions in the following sections. A more comprehensive overview of the light-cone gauge approach can be found in \cite{Ponomarev:2022vjb}.
	 
	\subsection{Free massless fields}
	In light-cone gauge the free action for a set of massless fields of helicities 
	$\lambda$  is
	\begin{equation}
		\label{1.6}
		S_2\equiv \int d^4 x L_2, \qquad L_2 = -\frac{1}{2}\sum_{\lambda}  \partial_a\Phi^{-\lambda} \partial^a \Phi^\lambda.
	\end{equation}
Lorentz invariance requires that opposite helicities enter in pairs, otherwise, the spectrum of a theory at the free level is arbitrary.

The action of the Poincare algebra in the helicity-$\l$ representation is given by
	\begin{eqnarray}
	\label{1.0}
			P^a \cdot \Phi^{\lambda}&\equiv& \partial^a \Phi^{\lambda},\\
			\label{1.1}
			J^{ab}\cdot \Phi^{\lambda}&\equiv& (x^a\partial^b - x^b \partial^a+S^{ab} )\Phi^\lambda,
	\end{eqnarray}
	where $S^{ab}$ is the spin part of the angular momentum. Its components act as 
	\begin{alignat}{2}
	\label{1.03}
			S^{+a}\cdot \Phi^{\lambda} &=0, &  S^{x\bar x} \cdot \Phi^{\lambda} &= -\lambda \Phi^{\lambda},\\
			S^{x-}\cdot \Phi^{\lambda} &= \lambda  \frac{\partial}{\partial^+}\Phi^{\lambda}, \qquad & 
			S^{\bar x -}\cdot \Phi^{\lambda}&= -\lambda   \frac{\bar\partial}{\partial^+}\Phi^{\lambda}.
			\label{1.3}
	\end{alignat}
	Our conventions on light-cone coordinates are given in  appendix \ref{Not}.
	
	The action (\ref{1.6}) is invariant with respect to the Poincare algebra transformations (\ref{1.0}), (\ref{1.1}). This invariance leads to the conservation of the corresponding Noether currents
	\begin{align}
		P^a\quad \to  \;\quad T^{a,b} &\,= \sum_{\lambda}\frac{\delta L_2}{\delta (\partial_b \Phi^\lambda)}\partial^a \Phi^\lambda - \eta^{ab} L_2,\\
		J^{ab} \quad   \to  \quad L^{ab,c} &\,= x^a T^{b,c}-x^b T^{a,c}+R^{ab,c},
		\label{1.12}
	\end{align}
	where $R^{ab,c}$ is the spin current
	\begin{equation}
		\label{1.13}
		R^{ab,c} \equiv \sum_{\lambda}\frac{\delta L_2}{\delta( \partial_k \Phi^\lambda)} S^{ab}\cdot \Phi^\lambda.
	\end{equation}

	The Noether charges are defined in the standard way
	\begin{equation}
		\label{1.14}
		Q_2[P^a] \equiv \int d^3x^\perp T^{a,+}, \qquad Q_2[J^{ab}] \equiv \int d^3x^\perp L^{ab,+}.
	\end{equation}
	Here the subscript "2" indicates that these charges are quadratic in fields and, thus, describe a free theory. Explicitly, these are given by 
	\begin{equation}
		\label{1.15}
		Q_2[P^i] = -\sum_{\lambda}\int d^3x^\perp \partial^+\Phi^{-\lambda}  p_2^i \Phi^\lambda, \qquad
		  Q_2[J^{ij}] = -\sum_{\lambda}\int d^3x^\perp \partial^+\Phi^{-\lambda}  j_2^{ij} \Phi^\lambda,
	\end{equation}
	where 
	\begin{align}
		\notag
		p_2^+ &\,=\partial^+, & p_2^- &= -\frac{\partial \bar\partial}{\partial^+}, \qquad\qquad \quad\;  p_2 = \partial,  \qquad\qquad\quad\;    \bar p_2=
		\bar\partial,\\
		\notag
		j_2^{+-} &\,=  - x^-\partial^+,&  j_2^{x\bar x} &= x\bar\partial - \bar x \partial -\lambda, \\
		\notag
		j_2^{x+}&\, = x\partial^+, &  j_2^{x-} &= -x\frac{\partial \bar\partial}{\partial^+} - x^- \partial+\lambda\frac{\partial}{\partial^+}, \\
		j_2^{\bar x+}&\, = \bar x\partial^+,
		&  j_2^{\bar x-} &= -\bar x\frac{\partial \bar\partial}{\partial^+} - x^- \bar\partial
		-\lambda\frac{\bar\partial}{\partial^+}.
		\label{1.16}
	\end{align}
In the light-cone approach it is conventional to choose $x^+$ as the time variable, while 
	 $x^{\perp}\equiv \{x,\bar x,x^- \}$ are regarded as spatial coordinates.
	 For our purposes it will be sufficient to consider charges (\ref{1.14}) at any fixed time, which,  for convenience, is chosen to be  $x^+=0$.
	
	The canonical momentum is defined in the standard way
	\begin{equation}
		\label{1jun2}
		\Pi^\lambda \equiv \frac{\delta L_2}{\delta( \partial^- \Phi^\lambda)}= -\partial^+ \Phi^{-\lambda}.
	\end{equation}
	Accordingly, the Hamiltonian reads
	\begin{equation}
		\label{1.8}
		H^\Phi_2\equiv  \sum_{\lambda }  \int  d^{3}x^\perp(\Pi^\lambda \partial^-\Phi^{\lambda} - L_2)  =
		\sum_{\lambda} \int d^{3}x^\perp \partial \Phi^{-\lambda} \bar\partial \Phi^\lambda,
	\end{equation}
	where  the integration is performed over equal-time hypersurfaces.
	
	Leaving out the analysis of constraints, we quote the resulting Dirac bracket
	\begin{equation}
		\label{1.9}
		[\partial^+ \Phi^\lambda(x^\perp,x^+),\Phi^{\mu}(y^\perp,x^+)]_\Phi= \frac{1}{2}\delta^{\lambda+\mu,0}\delta^{3}(x^\perp,y^\perp).
	\end{equation}
	To distinguish this Dirac bracket from the one that acts in the phase space of the point particle, we supplement the former with the subscript "$\Phi$".
	
By construction, charges (\ref{1.14}), once written in terms of fields and the associated momenta, generate Poincare transformations on the phase space of theory (\ref{1.6}) via the Dirac bracket (\ref{1.9}). This ensures that conserved charges associated with any pair of Poincare generators $O^{(1)}$ and $O^{(2)}$ commute as
	\begin{equation}
		[Q_2[O^{(1)}], Q_2[O^{(2)}]]_{\Phi}=Q_2[[O^{(1)},O^{(2)}]],
	\end{equation}
	where the commutator on the right-hand side is the Lie algebra commutator.

	\subsection{Free point particle}
	
	The action for a free point particle in Minkowski space is expressed as follows
	\begin{equation}
	\label{18feb2}
		S=-m\int d \tau \sqrt{-\eta_{ab}\dot x^{a} \dot x^{b}}=-m\int d \tau \sqrt{-2\dot x^{+}\dot x^- -2\dot x \dot{\bar x}},
	\end{equation}
	where $\dot x$ denotes the derivative with respect to the world-line parameter $\tau$ and $m$ denotes the mass of the particle. This action exhibits reparametrization invariance. To fix it, we set the world-line parameter to  the light-cone time  $\tau=x^+$. This  yields 
	\begin{equation}
		S=-m\int dx^{+}\sqrt{-2\dot x^- -2\dot x \dot{\bar x}}\equiv \int L_p dx^+.
		\label{action}
	\end{equation}

When proceeding with the Hamiltonian analysis, 
as in field theory, $x^{+}$ is interpreted as the time coordinate, while the remaining three coordinates $\{x,\bar x,x^- \}$ are treated as spatial coordinates.
The canonical momenta are defined in the standard way
\begin{equation}
\label{18deb1}
p_{\perp} \equiv \frac{\partial L_p}{\partial \dot{x}^{\perp}},
\end{equation}
which leads to the Hamiltonian
	\begin{equation}
	\label{18feb3}
		H_p=\frac{p_x p_{\bar x}}{p_-}+\frac{m^2 }{2p_-}.
	\end{equation}
The Poisson bracket is defined in the standard manner 
	\begin{equation}
		[f, g]_p\equiv \frac{\partial f}{\partial x^i}\frac{\partial g}{\partial p_i}-\frac{\partial f}{\partial p_i}\frac{\partial g}{\partial x^i},
	\end{equation}
	where the subscript "$p$" indicates that the bracket   acts on the coordinates and momenta associated with the point particle.

	The action (\ref{action}) exhibits Poincare invariance with the Poincare transformations given by
	\begin{equation}
		J^{ab}[x^c]=x^a\eta^{bc}-x^b \eta^{ac}, \qquad P^a[x^b]=\eta^{ab}.
		\label{inv}
	\end{equation}  
	In accordance with Noether's theorem, each Poincare generator is associated with a conserved quantity. These are explicitly given by
	\begin{equation}
		\label{1.28}
		\begin{split}
			q_0 [P^x] = -p^x, \qquad q_0 [P^{\bar x}] = -p^{\bar x}, \qquad q_0 [P^+] = -p^+, \qquad q_0 [P^-] = H_p,\\
			q_0 [J^{x\bar x}] = \bar xp^x-x p^{\bar x}, \qquad q_0 [J^{x+}] = -xp^+, \qquad 
			q_0 [J^{\bar x+}] = -\bar xp^+,\\
			q_0 [J^{x-}] = H_p x + p^x x^-, \qquad  q_0 [J^{\bar x-}] = H_p \bar x + p^{\bar x} x^-,
			\qquad  q_0 [J^{+-}] =  p^+ x^-.
		\end{split}
	\end{equation}
	As in the field theory case, we are free to consider the conserved quantities at any time, which was chosen as $x^+=0$, for simplicity.
	
	By construction, 
	conserved quantities (\ref{1.28}) form a representation of the Poincare algebra on the point-particle phase space. In other words, for any two generators $O^{(1)}$ and $O^{(2)}$  of the Poincare algebra, the following relation holds
	\begin{equation}
		[q_0[O^{(1)}], q_0[O^{(2)}]]_p=q_0[[O^{(1)},O^{(2)}]].
	\end{equation}
	
	At  later stages of the analysis of the present paper we will focus on the case of a massless particle. The above discussion may seem inapplicable to this case, as 
	the action (\ref{18feb2}) vanishes for $m=0$. The standard way to resolve this problem is to consider an alternative form of (\ref{18feb2}), which has a smooth massless limit at the expense of introducing an auxiliary  world-line metric. This alternative action requires a separate Hamiltonian analysis, which is carried out in appendix \ref{ApHam}.
As expected, we found  the same Hamiltonian,  as for the original action, (\ref{18feb3}). In other words, the naive massless limit of the above discussion is justified.

	\section{Constraints on interactions}
	\label{sec:3}
	
	In the previous section, we reviewed the description of the free massless higher-spin fields and the free point particle in Minkowski space using the light-cone approach. 
	In the present section, we will review how the light-cone formalism works at the interaction level. In particular,  we will show  how it constrains interactions of a point particle with chiral higher-spin fields. The associated constraints will be analysed in the following sections.

	\subsection{Generalities on interactions}

Once interactions are added, the free theory Poincare charges receive non-linear corrections. For the problem in question, the charges admit the following form
	\begin{equation}
	\begin{split}
		\label{3.1}
			\mathcal Q [O]& \equiv Q[O]+q[O], \\
			Q [O] &\equiv  Q_2[O]+Q_3[O], 
			\\
			q[O] &\equiv 
			q_0[O]+q_1[O]+q_2[O]+\dots.
	\end{split}
	\end{equation}
Here $\mathcal Q [O]$ is the total charge, associated with the generator $O$ and it has two contributions. The first contribution, $Q$, is the field theory term, which consists  of the free field theory contribution $Q_2$, followed by non-linear corrections. In general, the series of non-linear corrections can be infinite. In (\ref{3.1}) we kept only a single term due to the fact that chiral higher-spin theories have cubic interactions only. 
 In turn, $q$
is the point particle charge. It starts with the free point particle term $q_0$, followed by terms, which are responsible for interactions of a point particle with the fields. In particular, $q_1$ is linear in fields, $q_2$ is quadratic, etc.

The key consistency condition that is being imposed in the light-cone formalism is that the Poincare invariance of a theory should not be broken by interactions. As we mentioned above, in practice this is implemented by requiring that for any pair of generators $O^{(1)}$ and
	$O^{(2)}$  with 
	\begin{equation}
		[O^{(1)},O^{(2)}]=O^{(3)},
	\end{equation}
the corresponding Poincare charges must commute accordingly
	\begin{equation}
		[\Q^{(1)},\Q^{(2)}]=\Q^{(3)}.
			\label{3.3}
	\end{equation}
Here we used a shortcut notation $\Q^{(i)}\equiv \Q[O^{(i)}] $.

	Next, by decomposing both the charges and the Poisson bracket into contributions associated with the fields and the particle,  and taking into account that the field theory is invariant under Poincare transformations on its own,  (\ref{3.3}) leads to  
	\begin{equation}
		[q^{(1)},q^{(2)}]_p  +[q^{(1)},Q^{(2)}]_\Phi + [Q^{(1)},q^{(2)}]_\Phi = q^{(3)}.
		\label{3.2}
	\end{equation}
To arrive at (\ref{3.2}) we also removed the back-reaction terms. A detailed discussion of these terms  can be found in \cite{Ivanovskiy:2023aay}. Equations (\ref{3.2})  should be regarded as a set of consistency conditions on point-particle charges. The field theory charges are known and will be given  below. 

Constraints (\ref{3.2}) will be solved perturbatively. Namely, by plugging the perturbative expansion for charges (\ref{3.1}) into (\ref{3.2}) 
and collecting the terms that feature fields to the same power, one finds an infinite series of consistency conditions. These should then be solved 
for point-particle charges
order by order in powers of fields.

When interactions are introduced, only the charges that act transversely to $x^+=0$ need to be deformed \cite{Dirac:1949cp}.
The associated generators are
\begin{equation}
	 P^- ,\qquad  J^{x-}, \qquad  J^{\bar x-}.
\end{equation}
These are referred to as dynamical generators, while the remaining generators are called  kinematical. Thus, at the non-linear level one, schematically, has
\begin{equation}
	D =D_2 + \delta D, \qquad K = K_2,
\end{equation}
where $D$  and $K$ denote dynamical and kinematical generators respectively. 

Returning to the consistency conditions, one finds that these split into three types, depending on the types of generators they feature. The simplest type of the consistency conditions has the schematic form 
\begin{equation}
	[K,K]=K.
\end{equation}
These are satisfied automatically at the non-linear level, as the kinematical charges do not receive non-linear corrections. 
Next, one has the consistency conditions of the form
\begin{align}
	[K,D] = K \qquad &\Rightarrow \qquad [K,\delta D] = 0,\\
	[K,D] = D \qquad &\Rightarrow \qquad [K,\delta D] = \delta D.
\end{align}
Both of these constraints yield linear differential equations on the deformations  $\delta D$, which can be solved systematically at any order in fields. Constraints of this type are called kinematical. The last class of constraints has the form
\begin{equation}
	[D,D]=0.
\end{equation}
Constraints of this type are called dynamical. Due to the fact that dynamical constraints are quadratic in deformations, solving them presents the main difficulty of the light-cone formalism.

Once the solution to the consistency conditions is found, the point-particle action can be obtained by the standard formula
\begin{equation}
\label{28feb1}
S_p = \int  (p_x d{x}+ p_{\bar x} d{\bar x}+ p_- d{x}^- - q[P^-] dx^+).
\end{equation}

\subsection{Field theory generators}

Consistency conditions (\ref{3.2}) involve field theory generators. We will focus on the coupling of a point particle to the chiral higher-spin theory \cite{Metsaev:1991nb, Metsaev:1991mt, Ponomarev:2016lrm} and its Poisson limit \cite{Ponomarev:2017nrr}. Besides that, as a consistency check, we will also consider the well-understood case of a  coupling of a point particle to self-dual gravity  \cite{Chalmers:1996rq}\footnote{
Self-dual gravity is usually understood in terms of equations of motion only, which, besides that, feature only a single dynamical degree of freedom carrying helicity $+2$. To make this theory Lagrangian, while still maintaining Lorentz invariance, one needs to add the field of helicity $-2$. Here, when we are referring to self-dual gravity, we mean its Lagrangian version, which features both helicities.}. Below we will quote the deformations of the dynamical charges in all these three cases.

The aforementioned theories involve  corrections to charges, which are cubic in fields only. In all cases these can  be represented as
\begin{equation}
\label{19feb1}
\begin{split}
	&Q_3[P^{-}]=\sum\limits_{\l_1, \l_2,\l_3}\int d^{3+9}x^{\perp} \prod_{i=1}^3\delta^3(x^\perp-x^\perp_i) h_{\Phi}^{\l_1\l_2\l_3}
	\prod_{i=1}^3\Phi^{\lambda_i}(x^\perp_i),\\
	&Q_3[J^{x-}]=\sum\limits_{\l_1, \l_2,\l_3}\int d^{3+9}x^{\perp} \prod_{i=1}^3\delta^3(x^\perp-x^\perp_i) 
	\left(j^{\l_1\l_2\l_3}_{\Phi}+xh^{\l_1\l_2\l_3}_{\Phi}\right)\prod_{i=1}^3\Phi^{\lambda_i}(x^\perp_i),\\
	&Q_3[J^{\bar x-}]=\sum\limits_{\l_1, \l_2,\l_3}\int d^{3+9}x^{\perp} \prod_{i=1}^3\delta^3(x^\perp-x^\perp_i) 
	\left(\bar j^{\l_1\l_2\l_3}_{\Phi}+\bar xh^{\l_1\l_2\l_3}_{\Phi}\right)\prod_{i=1}^3\Phi^{\lambda_i}(x^\perp_i),
	\end{split}
\end{equation}
where 
\begin{equation}
\label{19feb1x1}
\begin{split}
	h_{\Phi}^{\l_1\l_2\l_3}&={C^{\l_1\l_2\l_3}}
	\frac{
		\bar{\mathbb{P}}
		^{\l_1+\l_2+\l_3}}
	{\partial^{+\l_1}_1\partial^{+\l_2}_2\partial^{+\l_3}_3},\\
	j^{\l_1\l_2\l_3}_{\Phi}&=
	\frac{2}{3} {C^{\l_1\l_2\l_3}} \left(\partial^{+}_1(\lambda_2-\lambda_3)+\partial^{+}_2(\lambda_3-\lambda_1)+\partial^{+}_3(\lambda_1-\lambda_2)\right)
	\frac{\bar{\mathbb{P}}^{\l_1+\l_2+\l_3-1}}
	{\partial^{+\l_1}_1\partial^{+\l_2}_2\partial^{+\l_3}_3},\\
	\bar j^{\l_1\l_2\l_3}_{\Phi}&=0	
	\end{split}
\end{equation}
and 
\begin{equation}
\label{28feb2}
\bar{\mathbb{P}}\equiv\frac{1}{3} \big(\partial_{x_3}(\partial^{+}_1-\partial^+_{2})+\partial_{x_1}(\partial^{+}_2-\partial^+_{3})+\partial_{x_2}(\partial^{+}_3-\partial^+_{1})\big).
\end{equation}
To be able to deal with differential operators acting on different fields efficiently, fields in (\ref{19feb1}) are initially put at different points. Once the differential operators have been applied, the result is evaluated at a single point, which in (\ref{19feb1})  is implemented by an explicit delta function. Finally, $C^{\l_1\l_2\l_3}$ are the coupling constants, which differ from theory to theory. 

For self-dual gravity only helicities $+2$ and $-2$ are present in the spectrum. The non-vanishing coupling constants are 
\begin{equation}
\label{28feb3}
C^{-2, 2, 2}=C^{2, -2, 2}=C^{2, 2, -2}=l,
\end{equation}
where $l$ is an arbitrary parameter with the dimension of length. For the chiral higher-spin theory all helicities are present in the spectrum and the coupling constants are 
\begin{equation}
\label{28feb4}
C^{\l_1\l_2\l_3}=\frac{l^{\l_1+\l_2+\l_3-1}}{\Gamma(\l_1+\l_2+\l_3)}.
\end{equation}
Finally, for the Poisson chiral higher-spin theory one has
\begin{equation}
\label{28feb5}
C^{\l_1\l_2\l_3}=2l\delta(\l_1+\l_2+\l_3-2).
\end{equation}

	\section{Kinematical constraints}
	\label{sec:4}

In this section we will start  solving the constraints on point-particle charges. 
First, we make the most general ansatz for dynamical charges at a given order in fields
\begin{eqnarray}
	&&q_n[P^-]=\sum\limits_{\lambda_i}\int d^{3n}z^\perp
	\left(\prod\limits_{i=1}^n\delta^3(x^\perp-z_i^{\perp})\right) h^{\lambda_1 \dots \lambda_n} \prod\limits_{i=1}^n \Phi^{\lambda_i}(z_i^{\perp}),
	\label{3.18}\\
	&&q_n[J^{x-}]=\sum\limits_{\lambda_i}\int d^{3n}z^\perp \left(\prod\limits_{i=1}^n\delta^3(x^\perp-z_i^{\perp})\right)
	 \left(
	j^{\lambda_1 \dots \lambda_n}
	+x \ h^{\lambda_1 \dots \lambda_n}\right)
	\prod\limits_{i=1}^n \Phi^{\lambda_i}(z_i^{\perp}),
	\label{3.19}\\
	&&q_n[J^{\bar x -}]=\sum\limits_{\lambda_i}\int d^{3n}z^\perp \left(\prod\limits_{i=1}^n\delta^3(x^\perp-z_i^{\perp})\right)\left(
	\bar j^{\lambda_1 \dots \lambda_n}
	+\bar x \ h^{\lambda_1 \dots \lambda_n}\right) \prod\limits_{i=1}^n \Phi^{\lambda_i}(z_i^{\perp}).
	\label{3.20}
\end{eqnarray}
Here $x$ is particle's coordinate, while $z$'s are coordinates of the fields. Functions $h^{\lambda_1 \dots, \lambda_n}$, $j^{\lambda_1 \dots \lambda_n}$ and $\bar j^{\lambda_1 \dots \lambda_n}$ are arbitrary functions of  $x^\perp$, $p^\perp$ and $\partial^\perp_{z_i}$.
 The presence of delta functions guarantees that the charges (\ref{3.18})-(\ref{3.20}) are local in the sense that the particle couples only to the values of fields and their derivatives at the space-time point  it is located.

We start by considering the kinematical constraints.
Those that involve the commutators with translations are easy to implement: they entail that $h^{\lambda_1 \dots \lambda_n}$, $j^{\lambda_1 \dots \lambda_n}$ and $\bar j^{\lambda_1 \dots \lambda_n}$ do not depend on coordinates $x^\perp$. 
The remaining kinematical constraints   lead to
\begin{align}
	\label{4.11}
	[J^{x+},P^-]=P^{x}\qquad \Rightarrow &\qquad \sum_{i=1}^n \frac{\partial h^{\lambda_1 \dots \lambda_n}}{\partial(\partial_{z_i})}\partial^+_i +p^+\frac{\partial}{\partial p_x}h^{\lambda_1 \dots \lambda_n}=0,\\
	\label{4.12}
	[J^{\bar x+},P^-]=P^{\bar x}\qquad \Rightarrow &\qquad \sum_{i=1}^n  \frac{\partial h^{\lambda_1 \dots \lambda_n}}{\partial(\partial_{\bar z_i})}\partial^+_i +p^+\frac{\partial}{\partial p_{\bar x}}h^{\lambda_1 \dots \lambda_n}=0,\\
	\label{4.13}
	[J^{+-},P^-]+P^-=0\qquad \Rightarrow &\qquad \sum_{i=1}^n \frac{\partial h^{\lambda_1 \dots \lambda_n}}{\partial (\partial^+_i)}\partial^+_i+p^+ \frac{\partial}{\partial p^+}h^{\lambda_1 \dots \lambda_n}+h^{\lambda_1 \dots \lambda_n}=0,\\
	[J^{x\bar x},P^-]=0\qquad \Rightarrow &\qquad \sum_{i=1}^n\left( 	\partial_{z_i} \frac{\partial h^{\lambda_1 \dots \lambda_n}}{\partial (\partial_{z_i})} 
	- \partial_{\bar z_i}\frac{\partial h^{\lambda_1 \dots \lambda_n}}{\partial (\partial_{\bar z_i})}\right)\nn
	&\qquad +\ p_x \frac{\partial h^{\lambda_1 \dots \lambda_n}}{\partial p_x} 
	- p_{\bar x}\frac{\partial h^{\lambda_1 \dots \lambda_n}}{\partial p_{\bar x}}-h^{\lambda_1 \dots \lambda_n}\sum_{i=1}^n\lambda_i =0.
	\label{4.14}	
\end{align}
The constraints on $j^{\lambda_1 \dots \lambda_n}$ are identical to those for $h^{\lambda_1 \dots \lambda_n}$, except that (\ref{4.14}) should be replaced with
\begin{align}
	[J^{x\bar x},J^{x-}]=J^{x-}\qquad \Rightarrow &\qquad \sum_{i=1}^n\left( 	\partial_{z_i} \frac{\partial j^{\lambda_1 \dots \lambda_n}}{\partial (\partial_{z_i})} 
	- \partial_{\bar z_i}\frac{\partial j^{\lambda_1 \dots \lambda_n}}{\partial (\partial_{\bar z_i})}\right)\nn
	&\qquad +p_x \frac{\partial j^{\lambda_1 \dots \lambda_n}}{\partial p_x} 
	- p_{\bar x}\frac{\partial j^{\lambda_1 \dots \lambda_n}}{\partial p_{\bar x}}-j^{\lambda_1 \dots \lambda_n}\sum_{i=1}^n\lambda_i+j^{\lambda_1 \dots \lambda_n} =0.
	\label{4.15}	
\end{align}
The constraints on $\bar j^{\lambda_1 \dots \lambda_n}$ are related to the constraints on $j^{^{\lambda_1 \dots \lambda_n}}$ by complex conjugation.

The first two conditions (\ref{4.11}), (\ref{4.12}) indicate that $h^{\lambda_1 \dots \lambda_n}$, $j^{\lambda_1 \dots \lambda_n}$ and $\bar j^{\lambda_1 \dots \lambda_n}$ can  only depend on $p_x, \partial_{z_i}$ and $p_{\bar x}, \partial_{\bar z_i}$   in combinations
\begin{equation}
	\sigma_{ z_i}=p_x-\partial_{z_i} \frac{p^+}{\partial^+_i}, \qquad \sigma_{ \bar z_i}=p_{\bar x}-\partial_{\bar z_i} \frac{p^+}{\partial^+_i}.
	\label{4.018}
\end{equation}
Together with  (\ref{4.13}) this means that $h^{\lambda_1 \dots \lambda_n}$, $j^{\lambda_1 \dots \lambda_n}$ and $\bar j^{\lambda_1 \dots \lambda_n}$ can be presented in the form 
\begin{equation}
\label{4.17}
\begin{split}
	h^{\lambda_1 \dots \lambda_n}&=\frac{A^{\lambda_1 \dots \lambda_n}(\sigma_{z_i}, \sigma_{\bar z_i}, s_i)}{p^+}, \\
		j^{\lambda_1 \dots \lambda_n}&=\frac{a^{\lambda_1 \dots \lambda_n}(\sigma_{z_i}, \sigma_{\bar z_i}, s_i)}{p^+},  \\
			\bar j^{\lambda_1 \dots \lambda_n}&=\frac{\bar a^{\lambda_1 \dots \lambda_n}(\sigma_{z_i}, \sigma_{\bar z_i}, s_i)}{p^+},
		\end{split}
\end{equation}
where
\begin{equation}
\label{28feb10}
s_i\equiv \frac{\partial^+_i}{p^+}
\end{equation}
and $A^{\lambda_1 \dots \lambda_n}$, $a^{\lambda_1 \dots \lambda_n}$ and $\bar{a}^{\lambda_1 \dots \lambda_n}$ are arbitrary functions of their arguments. Finally, (\ref{4.14}), (\ref{4.15}) and their $\bar{j}$ counterpart lead to
\begin{equation}
\begin{split}
	\sum_{i=1}^n \left(\sigma_{z_i}\frac{\partial A^{\lambda_1 \dots \lambda_n}}{\partial \sigma_{z_i}}-\sigma_{\bar z_i}\frac{\partial A^{\lambda_1 \dots \lambda_n}}{\partial \sigma_{\bar z_i}}-\lambda_i  A^{\lambda_1 \dots \lambda_n}\right)=0,\\
	\sum_{i=1}^n \left(\sigma_{z_i}\frac{\partial a^{\lambda_1 \dots \lambda_n}}{\partial \sigma_{z_i}}-\sigma_{\bar z_i}\frac{\partial a^{\lambda_1 \dots \lambda_n}}{\partial \sigma_{\bar z_i}}-\lambda_i  a^{\lambda_1 \dots \lambda_n}\right)+a^{\lambda_1 \dots \lambda_n}=0,\\
	\sum_{i=1}^n \left(\sigma_{z_i}\frac{\partial \bar a^{\lambda_1 \dots \lambda_n}}{\partial \sigma_{z_i}}-\sigma_{\bar z_i}\frac{\partial \bar a^{\lambda_1 \dots \lambda_n}}{\partial \sigma_{\bar z_i}}-\lambda_i  \bar a^{\lambda_1 \dots \lambda_n}\right)-\bar a^{\lambda_1 \dots \lambda_n}=0.
	\end{split}
\end{equation}

The kinematical constraints for point particles interacting with higher-spin fields were previously analysed for charges which are linear in fields \cite{Ivanovskiy:2023aay}. 
Above we extended these results to charges, which involve fields to  arbitrary powers.

\section{Dynamical constraints}
\label{sec:5}

There exist three dynamical constraints, which follow from  commutation relations
\begin{equation}
\label{20feb1}
  [P^-,J^{ x-}] = 0, \qquad [P^-,J^{\bar x-}]=0,\qquad  [J^{x-}, J^{\bar x -}]=0
\end{equation}
respectively. The third constraint is a consequence of the first two, as demonstrated in \cite{Ponomarev:2016lrm}. Therefore, it suffices to focus on solving only the first two constraints.

At the $n$-th order in fields the first constraint  gives
\begin{equation}
	\label{4.19}
\begin{split}
	[P^-,J^{x-}]=0 \qquad \Rightarrow \qquad &[q_n[P^-], q_0[J^{x-}]]_p+[q_{n-1}[P^-], q_1[J^{x-}]]_p+ \dots \\
	&+ [q_1[P^-], q_{n-1}[J^{x-}]]_p+ [q_0[P^-], q_n[J^{x-}]]_p\\
	&+[q_n[P^-], Q_2[J^{x-}]]_\Phi+[q_{n-1}[P^-], Q_3[J^{x-}]]_\Phi\\
	&+[Q_2[P^-], q_n[J^{x-}]]_\Phi+[Q_3[P^-], q_{n-1}[J^{x-}]]_\Phi=0.
	\end{split}
\end{equation}
To obtain (\ref{4.19}), we used (\ref{3.2}) for the given commutator, expanded the charges in powers of fields (\ref{3.1}) and collected the terms that involve fields to the $n$-th power.

When progressing with the perturbative analysis, every successive order of (\ref{4.19}) brings two new charges, which enter linearly. At the $n$-th order these are $q_n[P^-]$ and $q_n[J^{x-}]$, which are responsible for the $n$-th order interactions of a particle with fields. The remaining contributions involve lower-order charges quadratically. It is, thus, natural to rearrange (\ref{4.19}) as
\begin{equation}
\label{20feb2}
\begin{split}
&[q_n[P^-], q_0[J^{x-}]]_p + [q_n[P^-], Q_2[J^{x-}]]_\Phi\\
&\qquad\qquad+[q_0[P^-], q_n[J^{x-}]]_p+[Q_2[P^-], q_n[J^{x-}]]_\Phi\\
& \qquad\qquad\qquad\qquad= - [q_{n-1}[P^-], q_1[J^{x-}]]_p -\dots - [q_{1}[P^-], q_{n-1}[J^{x-}]]_p\\
&\qquad\qquad\qquad\qquad\qquad\qquad -[q_{n-1}[P^-], Q_3[J^{x-}]]_\Phi-[Q_3[P^-], q_{n-1}[J^{x-}]]_\Phi
\end{split}
\end{equation}
and regard it as a linear differential equation on $q_n[P^-]$ and $q_n[J^{x-}]$ with the right-hand side treated as an inhomogeneous source term.
By utilising (\ref{4.17}) and evaluating the left-hand side explicitly, we find
 \begin{equation}
 	\label{4.199}
 \begin{split}
 	&a^{\lambda_1 \dots \lambda_n}\sum\limits_{i=1}^n\left(\frac{m^2}{2}+\sigma_{z_i}\sigma_{\bar z_i}\right)s_i\\
 	&\qquad +\sum\limits_{i=1}^n\left(s_i\sigma_{\bar z_i}\frac{\partial }{\partial s_i}
 	-\sigma_{\bar z_i}^2 \frac{\partial}{\partial \sigma_{ \bar z_i}}
 	-\frac{m^2}{2}\frac{\partial}{\partial \sigma_{z_i}}-\l_i\sigma_{\bar z_i}
 	\right)A^{\lambda_1 \dots \lambda_n}=S^{\lambda_1\dots\lambda_n},
	\end{split}
 \end{equation}
where the source term  $S^{\lambda_1\dots\lambda_n}$ accounts for all contributions to the right-hand side of (\ref{20feb2}). The explicit expressions for the source terms are rather complicated. We will give them below at the relevant order.
 
Similarly, for the second constraint, $[P^-,J^{\bar x-}]=0$, one finds
 \begin{equation}
 \label{4.199x1}
 \begin{split}
	&\bar a^{\lambda_1 \dots \lambda_n}\sum\limits_{i=1}^n\left(\frac{m^2}{2}+\sigma_{z_i}\sigma_{\bar z_i}\right)s_i\\
	&\qquad +\sum\limits_{i=1}^n\left(s_i\sigma_{ z_i}\frac{\partial }{\partial s_i}
	-\sigma_{ z_i}^2 \frac{\partial}{\partial \sigma_{  z_i}}
	-\frac{m^2}{2}\frac{\partial}{\partial \sigma_{\bar z_i}}+\l_i\sigma_{ z_i}
	\right)A^{\lambda_1 \dots \lambda_n}=\bar S^{\lambda_1\dots\lambda_n}.
		\end{split}
\end{equation}

Before proceeding to the solution of the dynamical constraints, we need to comment on how locality is being imposed. By locality one, vaguely speaking, means, that interactions occur at a point. Mathematically, this translates into the requirement that interactions involve finitely many space-time derivatives or, probably, that higher-derivative terms are properly suppressed. In the light-cone formalism, as an artefact of the particular gauge fixing procedure, inverse powers of $\partial^+$ appear even for local theories. This can be seen already for the free theory charges (\ref{1.16}). Instead, infinite series or poles in transverse derivatives, $\partial_x$ and $\partial_{\bar x}$, do imply that the theory is non-local. Due to the kinematical constraints, these derivatives only appear in combinations $\sigma_x$ and $\sigma_{\bar x}$. Therefore, to impose locality, we need to constrain the dependence of charges on $\sigma_x$ and $\sigma_{\bar x}$ in some way. Considering that locality is a rather vague notion, 
justifying precise constraints on this dependence is a subtle issue. Luckily, to proceed with the problem we are dealing with, we do not need to do that. 
Indeed, at the technical level, the only relevant information about the analytic properties of charges needed is  whether these have poles at some specific locations, which, if present, trivialise the light-cone consistency conditions. For example,  by solving (\ref{4.199}) algebraically\footnote{Despite $\sigma$'s and $s$'s involve derivatives of fields, in (\ref{4.199}), (\ref{4.199x1}) these play the role of usual variables on which the unknown functions $A$, $a$ and $\bar a$ depend. Equations (\ref{4.199}) and (\ref{4.199x1}) should be then understood as differential equations for these functions. Since $a$ and $\bar a$, unlike $A$, contribute to (\ref{4.199}) and (\ref{4.199x1}) algebraically, one can easily solve for $a$ and $\bar a$ by dividing the both sides of (\ref{4.199}) and (\ref{4.199x1}) by their respective prefactors.} for $a^{\lambda_1 \dots \lambda_n}$, this constraint can be trivially satisfied. This will, however, lead to $a^{\lambda_1 \dots \lambda_n}$ with a pole in $\sigma_x$ and $\sigma_{\bar x}$ and, thus, this solution can be disregarded as non-local\footnote{In fact, this pole leads to an even more serious problem than non-locality: the charge diverges when the point particle and the field go on-shell.}. 
In the following, we will use locality only to remove such trivial and manifestly non-local solutions.

\subsection{Second-order consistency conditions}

	The solution of the dynamical constraints at the first order in fields was found in \cite{Ivanovskiy:2023aay}. The most general local corrections to the dynamical  generators of the point particle are expressed as 
	\begin{equation}
	\label{4.1}
	\begin{split}
		&h^{\l}= \left(C^{\lambda}\frac{\sigma^{\lambda}_x}{p^+}+\bar C^{-\lambda}\frac{\sigma^{-\lambda}_{\bar x}}{p^+}\right),\\
		&j^{\l}=\l C^{\l}\frac{\sigma^{\lambda-1}_{x}}{s_x p^+}, \qquad 	\bar j^{\l}=-\l \bar C^{-\l}\frac{\sigma_{\bar x}^{-\lambda-1}}{s_x p^+},
		\end{split}
	\end{equation}
	where $C^\lambda$ and $\bar C^{\lambda}$ are arbitrary coupling constants satisfying
	\begin{equation}
		\label{13jun4}
		C^{\lambda}=0, \qquad \lambda<0 \qquad \text{and} \qquad \bar C^{-\lambda}=0, \qquad \lambda \ge 0. 
	\end{equation}
	Due to the structure of equation (\ref{4.19}), interactions among higher-spin fields are irrelevant at this order. 

Next, we move on to the second-order analysis. The relevant equations to solve are (\ref{4.199}) and (\ref{4.199x1}). More explicitly, for the first equation one has
 \begin{equation}
  	\label{20feb5}
 \begin{split}
 	&a^{\lambda_1 \lambda_2}\left[\left(\frac{m^2}{2}+\sigma_{x}\sigma_{\bar x}\right)s_x+\left(\frac{m^2}{2}+\sigma_{y}\sigma_{\bar y}\right)s_y\right]\\
 	&\qquad+\left(s_x\sigma_{\bar x}\frac{\partial }{\partial s_x}
	+s_y\sigma_{\bar y}\frac{\partial }{\partial s_y}
 	-\sigma_{\bar x}^2 \frac{\partial}{\partial \sigma_{ \bar x}}
	-\sigma_{\bar y}^2 \frac{\partial}{\partial \sigma_{ \bar y}}\right.\\
	&\qquad\qquad \left.
 	-\frac{m^2}{2}\frac{\partial}{\partial \sigma_{x}}
	-\frac{m^2}{2}\frac{\partial}{\partial \sigma_{y}}
	-\l_1\sigma_{\bar x}-\l_2\sigma_{\bar y}
 	\right)A^{\lambda_1 \lambda_2}
	=S^{\lambda_1\lambda_2},
	\end{split}
 \end{equation}
where, compared to (\ref{4.199}), we changed notations,  $z_1^{\perp}\to x^{\perp}$, $z_2^{\perp}\to y^{\perp}$. To compute the source term on the right-hand side, we use the first order contribution to the charges of a point particle (\ref{4.1}), as well as the cubic charges on the field-theory side (\ref{19feb1})-(\ref{19feb1x1}).
Omitting a rather technical computation, we find 
\begin{equation}
		S^{\l_1\l_2}=S_1^{\l_1\l_2}+S_2^{\l_1\l_2}+S_3^{\l_1\l_2}+S_4^{\l_1\l_2}, 
	\end{equation}
	where
	\begin{equation}
		\label{20feb10}
	\begin{split}
		&S_1^{\l_1\l_2}=\frac{1}{2}C^{\lambda_1}C^{\lambda_2}\sigma_{x}^{\lambda_1}\sigma_{ y}^{\lambda_2-2}\lambda_2(\lambda_2-1)\frac{s_x(\sigma_{ x}-\sigma_y)}{ s_y}\\
		&\qquad\qquad \qquad\qquad \qquad\qquad+\frac{1}{2}C^{\lambda_1}C^{\lambda_2}\sigma_{y}^{\lambda_2}\sigma_{ x}^{\lambda_1-2}\lambda_1(\lambda_1-1)\frac{s_y(\sigma_{ y}-\sigma_x)}{ s_x},\\
		&S_2^{\l_1\l_2}=\bar C^{-\lambda_1} C^{\lambda_2}
		\sigma_{\bar x}^{-\lambda_1-1}\sigma_y^{\lambda_2-2}\Big[\lambda_1\lambda_2\sigma_{y}(\sigma_{\bar y}-\sigma_{\bar x})-\sigma_{\bar x}\frac{\lambda_2(\lambda_2-1)s_x}{s_y}(\sigma_{ y}-\sigma_x)\Big],
		\\
		&S_3^{\l_1\l_2}=\sum \limits_{\l}(-1)^{\l-1}\frac{3}{2} C^{\lambda} \ C^{-\lambda\lambda_1\lambda_2}\frac{
			(s_x\sigma_x+s_y \sigma_y)^{\l-1} (\sigma_y-\sigma_x)
			^{-\l+\l_1+\l_2-1}}
		{s^{\lambda-\lambda_2+1}_x s^{\l-\l_1+1}_y}\\
		&\qquad\qquad\qquad\qquad\qquad\qquad \qquad \Big[s_x\sigma_x\ (\l-\lambda_1+\lambda_2)+s_y \sigma_y (-\l-\lambda_1+\lambda_2)\Big],\\
		&S_4^{\l_1\l_2}=\sum \limits_{\lambda}(-1)^{\l+1}\frac{3}{2} \bar C^{-\l} \ C^{-\lambda\lambda_1\lambda_2}\frac{(\sigma_{\bar x} s_x+\sigma_{\bar y} s_y)^{-\l} (\sigma_y-\sigma_x)	^{-\lambda+\lambda_1+\lambda_2-1}}{s^{\lambda-\lambda_2+1}_x s^{\lambda-\lambda_1+1}_y(s_x+s_y)^{-2\l+1}}\\
		&\qquad\qquad\qquad\qquad\qquad\qquad\qquad \qquad\Big[s_x\ (\lambda-\lambda_1+\lambda_2)+s_y  (-\lambda-\lambda_1+\lambda_2)\Big].	
		\end{split}
	\end{equation}
Here  $S_1$ and $S_2$   arise from the commutators that involve only the point-particle charges. These are further split into two contributions depending on the signs of helicities of the fields involved. Sources $S_3$ and $S_4$, instead, originate from commutators that include both  the point-particle generators and the field-theory generators. Note that (\ref{20feb10}) does not involve negative powers of $\sigma$'s,  see (\ref{13jun4}).

Similarly, we consider the second constraint. At the second order in fields one finds
 \begin{equation}
  	\label{22feb3}
 \begin{split}
 	&\bar{a}^{\lambda_1 \lambda_2}\left[\left(\frac{m^2}{2}+\sigma_{x}\sigma_{\bar x}\right)s_x+\left(\frac{m^2}{2}+\sigma_{y}\sigma_{\bar y}\right)s_y\right]\\\
 	&\qquad+\left(s_x\sigma_{ x}\frac{\partial }{\partial s_x}
	+s_y\sigma_{ y}\frac{\partial }{\partial s_y}
 	-\sigma_{ x}^2 \frac{\partial}{\partial \sigma_{  x}}
	-\sigma_{y}^2 \frac{\partial}{\partial \sigma_{  y}}\right.\\
	&\qquad\qquad \left.
 	-\frac{m^2}{2}\frac{\partial}{\partial \sigma_{\bar x}}
	-\frac{m^2}{2}\frac{\partial}{\partial \sigma_{ \bar y}}
	+\l_1\sigma_{ x}+\l_2\sigma_{ y}
 	\right)A^{\lambda_1 \lambda_2}
	=\bar{S}^{\lambda_1\lambda_2},
	\end{split}
 \end{equation}
where the source term is
\begin{equation}
\bar	S^{\l_1\l_2}=\bar S_1^{\l_1\l_2}+\bar S_2^{\l_1\l_2}+\bar S_3^{\l_1\l_2}
\end{equation}
and
	\begin{equation}
		\label{22feb4}
	\begin{split}
	\bar S_1^{\l_1\l_2}&=-\frac{1}{2}\bar C^{-\lambda_1}\bar C^{-\lambda_2}\sigma_{\bar x}^{-\lambda_1}\sigma_{\bar  y}^{-\lambda_2-2}\lambda_2(\lambda_2+1)\frac{s_x(\sigma_{\bar  x}-\sigma_{\bar y})}{ s_y}\\
	&	\qquad\qquad\qquad\qquad\qquad\qquad-\frac{1}{2}\bar C^{-\lambda_1}\bar C^{-\lambda_2}\sigma_{\bar x}^{-\lambda_2}\sigma_{\bar  y}^{-\lambda_1-2}\lambda_1(\lambda_1+1)\frac{s_x(\sigma_{\bar  y}-\sigma_{\bar x})}{ s_y},\\
	\bar S_2^{\l_1\l_2}&=C^{\lambda_1} \bar C^{-\lambda_2}
	\sigma_{ x}^{\lambda_1-1}\sigma_{\bar y }^{-\lambda_2-2}\Big[\lambda_1\lambda_2\sigma_{\bar y}(\sigma_{ x}-\sigma_{ y})+\sigma_{  x}\frac{\lambda_2(\lambda_2+1)s_x}{s_y}(\sigma_{\bar  x}-\sigma_{\bar y})\Big],\\
	\bar S_3^{\l_1\l_2}&=\sum \limits_{\l}\l (-1)^{\l-1}3 \bar C^{-\l} \ C^{-\lambda\lambda_1\lambda_2}\frac{
		(s_x\sigma_{\bar x}+s_y \sigma_{\bar y})^{-\l-1}(s_x+s_y)^{2\l-1} (\sigma_y-\sigma_x)
		^{-\l+\l_1+\l_2}}
	{s^{\l-\l_2}_x s^{\l-\l_1}_y}.
	\end{split}
\end{equation}

\subsection{Analysis of the massless case}

In the present paper, we will focus on solving  (\ref{20feb5}) and (\ref{22feb4}) in the special case of 
\begin{equation}
\label{20febx10}
m=0.
\end{equation} 
In this case, the analysis of equations simplifies substantially thanks to a trick, which is analogous to the one used in the field theory case \cite{Metsaev:1991nb, Metsaev:1991mt}. Namely, one observes that the prefactor of  $a$ and the differential operator acting on $A$ in (\ref{20feb5}) both vanish for 
\begin{equation}
\label{20feb11}
\sigma_{\bar x}=\sigma_{\bar y}=0.
\end{equation}
 Considering that by our locality assumption $a^{\l_1\l_2}$ and $A^{\l_1\l_2}$ are free of poles in transverse derivatives, the left-hand side of (\ref{20feb5}) vanishes at (\ref{20feb11}).
 This means that (\ref{20feb5}) entails 
 	\begin{equation}
		S^{\l_1\l_2}|_{\sigma_{\bar x}=\sigma_{\bar y}=0}=0.
		\label{20feb12}
	\end{equation}
By examining the structure of individual source contributions  (\ref{20feb10})  it is trivial to see that (\ref{20feb12}) is equivalent to
\begin{equation}
		S_1^{\l_1\l_2}+S_3^{\l_1\l_2}=0.
		\label{4.6}
	\end{equation}
	
The major advantage of (\ref{4.6}) compared to the original equation (\ref{20feb5}) is that the former does not involve the second order charges, represented by unknown functions $a^{\lambda_1\lambda_2}$ and $A^{\lambda_1\lambda_2}$. Instead, it only involves the first order charges, which, moreover, have been already fixed by the first order analysis up to 
the coupling constants $C^{\l}$ and $\bar C^{-\l}$. In other words, instead of dealing with differential equations involving two unknown functions, by moving to (\ref{4.6}), we reduced the problem to a closed system of equations on the first-order coupling constants.

Equation (\ref{22feb3}) in the massless case can be treated similarly. Namely, one notices that for 
\begin{equation}
\label{22feb6}
\sigma_{ x}=\sigma_{ y}=0
\end{equation}
its left-hand side vanishes. As a result, one finds
\begin{equation}
\bar	S^{\l_1\l_2}|_{\sigma_{x}=\sigma_{ y}=0}=0.
\label{B10}
\end{equation}
By inspecting the explicit form of source terms in (\ref{22feb4}), one finds that (\ref{B10}) is equivalent to 
\begin{equation}
	\bar S_1^{\l_1\l_2}=0.
	\label{4.9}
\end{equation}

Equation (\ref{4.9}) is a functional equation for unknown $\bar C$'s, that needs to be satisfied for any pair $\lambda_1$ and $\lambda_2$. Due to the fact that $\bar S_1^{\l_1\l_2}$ contains two terms with different powers of $\sigma_{\bar x}$ and $\sigma_{\bar y}$, it can only vanish if they both come with the vanishing prefactors. Therefore,
	\begin{equation}
	\label{22feb7}
		\bar C^{-\lambda}=0, \qquad \forall \lambda < 0.
	\end{equation}
In other words, we find that  a massless point particle cannot couple to negative helicity fields of chiral theories. 

The analysis of the point-particle coupling to positive-helicity fields is a bit more complicated. Below, we will solve (\ref{4.6}) in three cases: self-dual gravity, the chiral higher-spin theory and its Poisson limit.

\paragraph{Self-dual gravity.}
As a first example, we will consider the case of self-dual gravity. Interactions of point particles with self-dual gravity are well understood. Below, we will demonstrate how these are reproduced in the light-cone gauge formalism.

It is not hard to see that (\ref{4.6}) is trivially satisfied unless both helicities  equal $+2$. In the letter case, one finds
	\begin{eqnarray}
	\notag
		&&(C^{2})^2(\sigma_{x}-\sigma_y)\left(\sigma_{y}^{2}\frac{s_y}{ s_x}-\sigma_{x}^{2}\frac{s_x}{ s_y}\right)\\
		&&\qquad\qquad\qquad\qquad-
		3 C^{2}l\frac{
			(s_x\sigma_x+s_y \sigma_y) (\sigma_x-\sigma_y)}
		{s_x s_y}(s_x\sigma_x\ -s_y \sigma_y )=0.
	\end{eqnarray}
	It leads to
	\begin{equation}
		C^{2}(C^{2}+3l)=0.
	\end{equation}
Therefore, Lorentz invariance requires that point particles are either free or interact with self-dual gravity with the coupling constant
		\begin{equation}
		\label{21feb1}
		C^{2}=-3l.
	\end{equation}
	The resulting action agrees with the standard covariant one, written in the Hamiltonian form and with the light-cone gauge imposed \cite{Ivanovskiy:2024ads}.
Let us also remark that (\ref{21feb1}) states that a point particle interacts with gravity with the coupling constant, which,
up to an inessential notation-dependent factor,
 is equal to Newton's gravitational constant. This result can be regarded as a point-particle version of Weinberg's equivalence principle \cite{Weinberg:1964ew}, stating that all fields interact with gravity with Newton's gravitational constant.

	\paragraph{Poisson chiral  higher-spin theory.} Next, we consider the case of the Poisson chiral higher-spin theory (\ref{28feb5}).
Consistency condition (\ref{4.6}) has to be satisfied for all pairs of helicities and all values of variables, that have not been set to zero in (\ref{20feb11}).
Considering (\ref{4.6}) for $\l_1\geq2$ and $\l_2=0$, we find 
	\begin{equation}
	\begin{split}
	\label{22feb2}
		&\frac{1}{2}C^{\lambda_1}C^{0}\sigma_{ x}^{\lambda_1-2}\lambda_1(\lambda_1-1)\frac{s_y(\sigma_{ x}-\sigma_y)}{ s_x}\\
		&\qquad \qquad\qquad=
		3 l C^{\lambda_1-2}\frac{
			(s_x\sigma_x+s_y \sigma_y)^{\l_1-3} (\sigma_y-\sigma_x)s_y}
		{s^{\lambda_1-1}_x } \Big[s_x\sigma_x\ +s_y \sigma_y (\l_1-1)\Big].
		\end{split}
	\end{equation}
After dividing the both sides by $(\sigma_{ x}-\sigma_y)$, one finds that the left-hand side, unlike the right-hand side, does not depend on $\sigma_y$. Thus, functions on the left-hand  and on the right-hand sides of (\ref{22feb2}) are linearly independent and the equality can be achieved only when both sides are vanishing. This leads to 
	\begin{equation}
		C^{\lambda}=0,\qquad  \forall \lambda \ge 0.
		\label{4.10}
	\end{equation}
Together with (\ref{22feb7}), this implies that a massless point particle cannot consistently couple to the Poisson chiral higher-spin theory.

\paragraph{Chiral higher-spin theory.} The analysis in the chiral higher-spin theory case proceeds in a similar manner. Namely, we consider (\ref{4.6}) for $\l_1\geq2$ and $\l_2=0$,
which leads to
	\begin{equation}
	\begin{split}
	&\frac{1}{2} C^{\lambda_1}C^{0}\sigma_{ x}^{\lambda_1-2}\lambda_1(\lambda_1-1)\frac{s_y(\sigma_{ x}-\sigma_y)}{ s_x}\\
	&\qquad\qquad\qquad	=-\frac{3}{2}\sum\limits_{\l} \frac{ C^{\lambda}\ l^{\l_1-\l-1}}{\Gamma(\l_1-\l)} \frac{
			(s_x\sigma_x+s_y \sigma_y)^{\l-1} (\sigma_y-\sigma_x)
			^{-\l+\l_1-1}}
		{s^{\lambda+1}_x s^{\l-\l_1+1}_y}\\
	&\qquad\qquad\qquad\qquad\qquad \qquad\qquad\qquad \quad \	 \Big[s_x\sigma_x\ (\l-\lambda_1)-s_y \sigma_y (\l+\lambda_1)\Big].
		 \end{split}
	\end{equation}
It is not hard to see, that it cannot be satisfied non-trivially either. For example, one can notice that the total homogeneity degree of the left-hand side in $s_x$ and $s_y$ is zero. At the same time, the right-hand side contains a series of terms with the total homogeneity degree  of the $C^\lambda$ term being $\lambda_1-\lambda-2$. This, in turn, implies that all $C^\lambda$ should be vanishing, unless $\lambda=\lambda_1-2$. Considering that this should be true for any $\lambda_1\ge 2$, we conclude that 
		\begin{equation}
		C^{\lambda}=0,\qquad  \forall \lambda\geq 0,
	\end{equation}
that is a massless point particle cannot couple to the chiral higher-spin theory either.

\section{Conclusion}
\label{sec:6}

In the present paper we studied interactions of point particles with chiral higher-spin theories using the light-cone formalism. By analysing the consistency conditions at the second order in interactions, we found that no local interactions of scalar point particles with chiral higher-spin theories are possible. This conclusion applies both to the chiral higher-spin theory and to its Poisson limit. As a consistency test of our approach, we checked that point particles can, indeed, interact with self-dual gravity, reproducing the standard result. 
We focused on a technically simpler case of massless point particles. The massive case requires a separate analysis, which will be presented elsewhere. Considering the fact that chiral higher-spin theories form inevitable closed subsectors of any massless higher-spin theory which admits Minkowski space as a solution, we expect that our results hold more generally, in particular, they apply to putative parity-invariant completions of chiral higher-spin theories. 

As we argued in the introduction, the possibility to couple a scalar point particle to higher-spin fields is closely tied to the existence of higher-spin geometry. Our results indicate that  Riemannian geometry cannot be extended to spaces with massless higher-spin fields. In particular, given a stationary spherically symmetric solution of higher-spin equations of motion, one will not be able to consistently locate its horizon or singularity, at least, unless these notions are substantially modified. Similarly, the geometric approach to conformal higher-spin theories \cite{Segal:2002gd} cannot lead to massless higher-spin theories, unless it is  substantially changed. Let us remark,  that this situation is in stark contrast with that in General Relativity, which can be consistently embedded into conformal gravity, see \cite{Maldacena:2011mk}, moreover, point particles couple to both theories in the same way\footnote{Note, however, that the point-particle coupling breaks the Weyl invariance of conformal gravity. If the Weyl invariance is kept, Riemannian geometry gets replaced with conformal geometry.}. For previous discussions on the embedding of massless higher-spin theories into conformal higher-spin theories, see \cite{Beccaria:2016syk,Adamo:2018srx}.

Our result applies to scalar point particles, which are idealised objects even in classical physics. These are designed to mimic non-spinning macroscopic bodies in problems for which their size and the internal structure can be neglected. Our result does not rule out consistent couplings of classical non-spinning macroscopic bodies to massless higher-spin fields, though, it implies that these descriptions do not have a consistent point-particle limit or, equivalently,  some information about the macroscopic structure of a body needs to be retained in any consistent limit. In particular, the proper time, measured by such a body in a higher-spin background should depend not only on its trajectory alone, but on a trajectory in a space, extended with additional parameters.

Let us note, that inconsistency of interactions of point particles with chiral higher-spin fields,  to some extent, could have been anticipated. Namely, the standard consistency requirement in quantum field theory states that fields should transform as unitary representations of the global symmetry algebra of a theory. Its point-particle counterpart states that a symmetry algebra should admit a realisation in terms of symplectic transformations on point-particle phase space \cite{Souriau}\footnote{For some recent developments in this area as well as for a more comprehensive list of references we refer the reader to \cite{Figueroa-OFarrill:2023vbj,Basile:2023vyg}.}. This means, that 
in order for a consistent coupling between a point particle and the chiral higher-spin theory to exist, it is necessary that the global chiral higher-spin algebra \cite{Ponomarev:2017nrr,Krasnov:2021nsq,Sharapov:2022faa,Ponomarev:2022atv} admits a realisation in terms of symplectic transformations on a phase space of a point particle. Despite, this problem has not been studied systematically before, closely related results suggest that this should be impossible. In particular, the global higher-spin algebra of massless theories in anti-de Sitter space \cite{Fradkin:1986ka,Eastwood:2002su,Vasiliev:2003ev} does not admit bulk scalar fields as representations. This is consistent with the fact that a single scalar field in anti-de Sitter space cannot couple to the higher-spin theory. It seems reasonable to expect, that the same applies to the coupling of scalar point particles and that this conclusion carries over to flat space as well\footnote{Validity of this argument can be further illustrated by other higher-spin examples. Namely, the higher-spin algebra in anti-de Sitter space admits the boundary conformal scalar field as a representation, which is consistent with the fact that conformal higher-spin theories on the boundary, having the same global symmetry, can couple to the boundary conformal scalar field. Moreover, the classical limit of conformal higher-spin theory, which amounts to the replacement of the Moyal bracket with the Poisson one, can couple to a point particle, that is the classical limit of a scalar field.}.

This line of reasoning not only allows one to rule out potential couplings, but also suggests types of particles that can couple to chiral higher-spin theories. In particular, the chiral higher-spin algebra admits an infinite tower of massless fields as a representation. This suggests that it should be possible to couple the classical limit of the chiral higher-spin theory -- presumably, its Poisson limit -- to an infinite tower of spinning massless particles. An explicit lower-spin implementation of such a limiting procedure at the level of action can be found in \cite{Ivanovskiy:2024ads}. In this regard, let us  note, that the generation of couplings of point particles to field theories via the proper classical limit seems to be a far more efficient way of finding these couplings, compared to the full-fledged analysis within the light-cone formalism carried out in the present paper.
Let us also note, that this logic suggests that chiral higher-spin theories should admit a coupling to flat-space singletons \cite{Ponomarev:2022ryp,Ponomarev:2022qkx}.
 It would be interesting to explore these ideas in future.

\acknowledgments

We would like to thank M. Grigoriev, E. Joung, S. Prohazka and E. Skvortsov  for interesting discussions on the interpretation of our results. We would also like to thank E. Skvortsov for comments on the draft.

\appendix
	\section{Notations \label{Not}}

We use the mostly plus convention for  the 4d Minkowski metric
\begin{equation}
	ds^2 = -(dx^0)^2+ (dx^1)^2+(dx^2)^2+(dx^3)^2.
\end{equation}
The light-cone coordinates are defined by
\begin{align}
	\notag
	x^+ &\,= \frac{1}{\sqrt{2}}(x^3+x^0), & x^-&\, = \frac{1}{\sqrt{2}}(x^3-x^0),\\
	\label{29sep1}
	x &\,=\frac{1}{\sqrt{2}}(x^1-ix^2), & \bar x &\,= \frac{1}{\sqrt{2}}(x^1+ix^2).
\end{align}
In these coordinates the metric reads
\begin{equation}
	ds^2 = 2dx^+ dx^- + 2 dx d\bar x.
\end{equation}
The coordinate transformation (\ref{29sep1}) leads to
\begin{align}
	\notag
	\partial^-  &\,= \frac{1}{\sqrt{2}}(\partial^3-\partial^0), &  \partial^+ &\, = \frac{1}{\sqrt{2}}(\partial^3 + \partial^0),\\
	\bar\partial &\,= \frac{1}{\sqrt{2}}(\partial^1-i\partial^2), &  \partial &\, =\frac{1}{\sqrt{2}}(\partial^1 + i \partial^2).
\end{align}
In particular,
\begin{equation}
	\partial^+ x^- = \partial^- x^+ = \bar\partial x = \partial\bar x = 1.
\end{equation}

\section{Point particle Hamiltonian with world-line metric} \label{ApHam}

In this appendix we revisit the free point particle in the massless case. To this end, we use the alternative action, which has the smooth massless limit. Our goal is to show that the Hamiltonian analysis of this action leads to the same results as in the main text, thus, justifying their massless limit.

	Consider the action for a free point particle
\begin{equation}
	S=-m\int dx^{+}\sqrt{-2\dot x^- -2\dot x \dot{\bar x}}.
\end{equation}
This action becomes singular for  $m=0$. To resolve this issue, one introduces an auxiliary field  $e$
\begin{equation}
	S=-\frac{1}{2}\int d \tau\left[ \frac{-2\dot x^- -2\dot x \dot{\bar x}}{e}+{m^2}e\right].
	\label{action2}
\end{equation}
This formulation of the action is well-defined even when  $m=0$.

The canonical momenta for the action (\ref{action2}) are defined in the standard manner
\begin{equation}
	\begin{split}
		p_-\equiv \frac{\partial L}{\partial \dot x^-}=\frac{1}{e},\\
		p_x\equiv \frac{\partial L}{\partial \dot x}=\frac{\dot{\bar{x}}}{e},\\
		p_{\bar x}\equiv \frac{\partial L}{\partial \dot \bar x}=\frac{\dot x}{e},\\
		p_{e}\equiv \frac{\partial L}{\partial \dot e}=0.
	\end{split}
	\label{momenta}
\end{equation} 
The canonical Hamiltonian for the point particle can then be expressed as
\begin{equation}
	\label{Ham2}
	H\equiv p_-\dot x^-+p_{x}\dot x + p_{\bar x}\dot{\bar{x}}+p_e \dot e-L = \left(p_x p_{\bar x}+\frac{m^2 }{2}\right)e.
\end{equation}

Equations (\ref{momenta})
lead to two primary constraints
\begin{eqnarray}
	ep_--1=0, \qquad p_e=0.
	\label{constr}
\end{eqnarray}
Following Dirac's procedure, we add these constraints to the Hamiltonian with arbitrary coefficients  $\a$, $\b$
\begin{equation}
	H_p  = \left(p_x p_{\bar x}+\frac{m^2 }{2}\right)e + \a p_e+\b(p_-e-1).
	\label{Hf}
\end{equation}
To determine the coefficients  $\a$, $\b$ , it is necessary to set the Poisson brackets of the complete Hamiltonian (\ref{Hf}) with the constraints specified in equation (\ref{constr}) to zero. This leads to the following conditions
\begin{eqnarray}
	&&	[H_f, p_e]_p=\left(p_x p_{\bar x}+\frac{m^2 }{2}\right)+\beta p_-=0,\nn
	&&	[H_f, p_-e-1]_p=\a=0.
\end{eqnarray} 
From these equations, we find
\begin{equation}
	\a=0, \qquad \b=-\frac{p_x p_{\bar x}}{p_-}-\frac{m^2 }{2p_-}.
	\label{coef}
\end{equation}
Substituting these expressions back into equation (\ref{Hf}), we obtain
\begin{equation}
	H_p=\frac{p_x p_{\bar x}}{p_-}+\frac{m^2 }{2p_-}.
\end{equation}
As expected, this results in the same Hamiltonian as we found in the main text.

\bibliography{pp}
\bibliographystyle{JHEP}

\end{document}